\documentclass[twocolumn]{revtex4-1}

\usepackage[utf8]{inputenc} 
\usepackage{amsmath}
\usepackage{amssymb}
\usepackage[normalem]{ulem}
\usepackage[margin=1in]{geometry}
\usepackage{xcolor}
\usepackage{siunitx}

\usepackage{graphicx}
\usepackage{dcolumn}
\usepackage{bm}

\begin{document}
	
	\title{Linear and angular motion of self-diffusiophoretic Janus particles}
	
	\author{Jérôme Burelbach}
	\email{de\_burel@hotmail.com}
	\author{Holger Stark}
	\email{holger.stark@tu-berlin.de}
	\affiliation{Institut f\"ur Theoretische Physik, Technische Universit\"at Berlin, Hardenbergstraße 36, 10623 Berlin, Germany}%
	
	\date{\today}
	
	\begin{abstract}
		We theoretically study the active motion of self-diffusiophoretic Janus particles (JPs) using the Onsager-Casimir reciprocal relations. The linear and angular velocity of a single JP are shown to respectively result from a coupling of electrochemical forces to the fluid flow fields induced by a force and torque on the JP. A model calculation is provided for half-capped JPs catalysing a chemical reaction of solutes at their surface, by reducing the continuity equations of the reacting solutes to Poisson equations for the corresponding electrochemical fields. We find that an anisotropic chemical activity alone is enough to give rise to active linear motion of a JP, whereas active rotation only occurs if the JP is not axisymmetric. In the absence of specific interactions with the solutes, the active linear velocity of the JP is shown to be related to the stoichiometrically weighted sum of the friction coefficients (or hydrodynamic radii) of the reacting solutes. Our reciprocal treatment further suggests that a specific interaction with the solutes is required to observe far-field diffusiophoretic interactions between JPs, which rely on an interfacial solute excess at the JP surface. Most notably, our approach applies beyond the boundary-layer approximation and accounts for both the diffusio- and electrophoretic nature of active motion.
	\end{abstract}
	
	\maketitle
	
	\section{Introduction}
	
	Active suspensions display an intricate collective behaviour that finds its practical use in a wide range of biophysical applications, such as helical swimming, \cite{Jekely2008} dynamic clustering, \cite{theurkauff2012,palacci2013}
	or self-assembled micromotors. \cite{Maggi2015} Recently, the active motion of self-phoretic Janus particles (JPs) has been reproduced successfully by a set of phenomenological Langevin equations, and the observed phase behaviour has been verified by a stability analysis using a generalised Keller-Segel model. \cite{Pohl2014,Liebchen2017,stark2018} 
	This analysis necessarily raises the question to what extent the phenomenological coefficients in these equations might be related to each other. For instance, a reduction to only two dimensionless parameters has been achieved for half-capped particles in 2D,\cite{Liebchen2017} based on earlier theoretical work on active motion within the boundary-layer approximation.\cite{Golestanian2007,Bickel2014} In brief, these theoretical models use an analogy to conventional phoretic motion in order to relate the velocity of a particle to its phoretic surface mobility, which for chemically passive particles relies on a specifc interaction with the surrounding fluid medium.\cite{Anderson1985} However, it has been noted that self-diffusiophoretic motion differs from diffusiophoresis of passive particles in that there is no solvent back-flow in the bulk of the system,\cite{Brady2011} thus making a direct analogy questionable. This suggests that further work is required to completely elucidate the physical nature of the transport coefficients that govern active motion of self-diffusiophoretic JPs.
	
	Here, we use the Onsager-Casimir reciprocal relations \cite{Onsager1931,Onsager1931a,Casimir1945} to formulate a general description of the linear and angular motion of self-diffusiophoretic JPs. Our description applies beyond the boundary-layer approximation and suggests that active self-diffusiophoretic motion can persist in the absence of specific interactions with the fluid, provided that the chemical activity of the JPs is anisotropic. A far-field model for diffusiophoretic interactions is then derived by noting that the solvent maintains a hydrostatic equilibrium relatively far away from the JP surface. Finally, the resulting expressions are evaluated for half-capped JPs that catalyse a chemical reaction at their surface.
	
	\section{Diffusiophoretic motion inside active suspensions: General Theory}
	
	\subsection{Active motion of single self-diffusiophoretic Janus particles\label{sec:-1}}
\begin{figure}
		\centering{}\includegraphics[width=3.8cm,height=3.6cm]{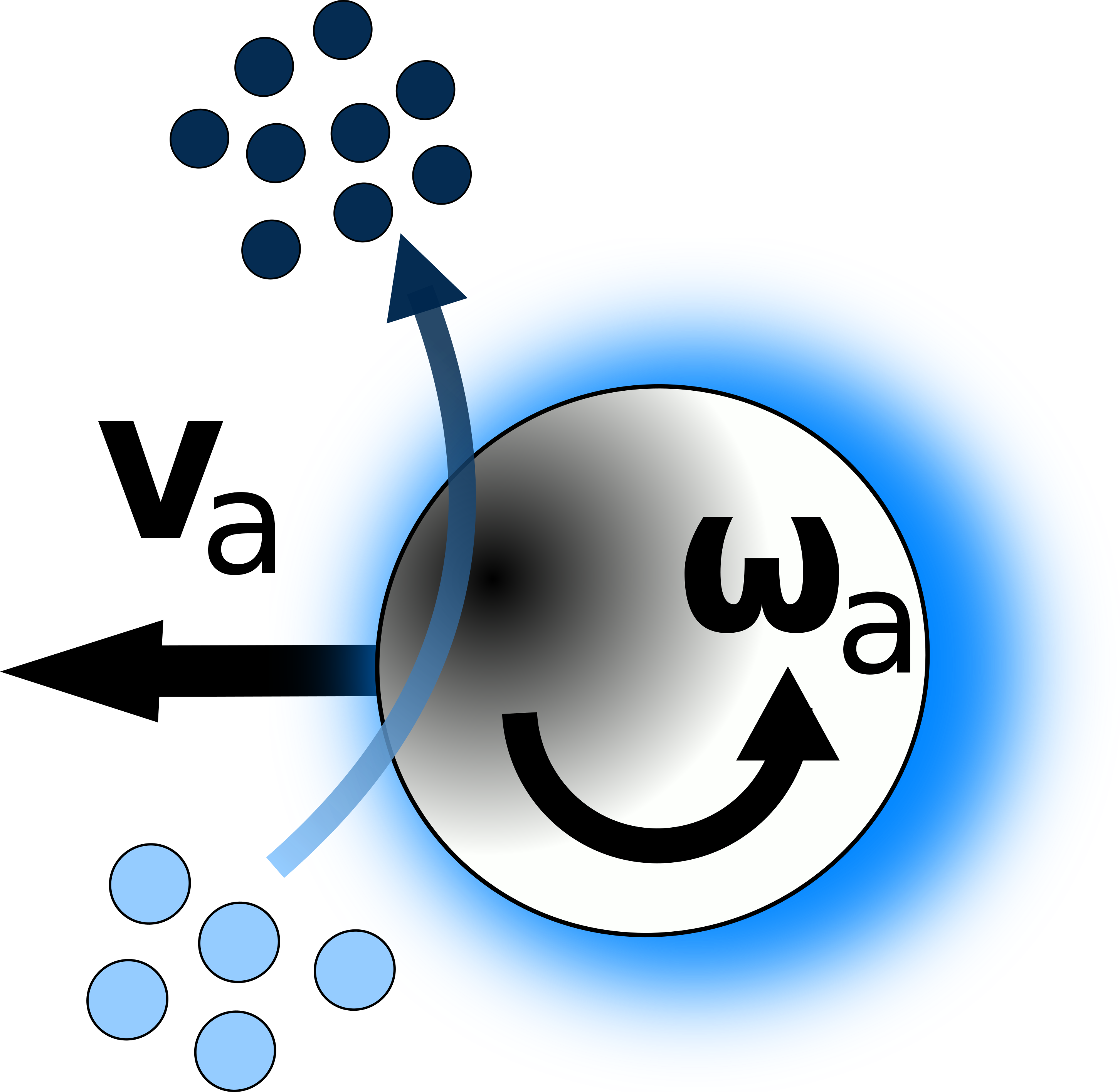}\caption{A JP catalysing a chemical reaction of a solutes at its surface. The solutes (light blue spheres) react with the chemically active part of the JP surface (shaded area), yielding a product with a different chemical composition (dark blue spheres). Moreover, the specific interaction between the solutes and the JP surface gives rise to an interfacial solute layer around the JP (blue radial gradient).}\label{fig:-1}
	\end{figure}

	We consider a self-diffusiophoretic JP with a hydrodynamic radius $R$, immersed in a fluid within a volume element $V$ at local thermodynamic equilibrium (LTE). The fluid mainly consists of an incompressible, viscous solvent, and can additionally contain several charged or uncharged solutes. Here, we will use the index $i$ for the fluid components, with $i=0$ referring to the solvent, and $i\neq 0$ to the solutes. The solutes may undergo a chemical reaction at the surface of the JP, thereby creating non-uniform electrochemical fields $\tilde{\mu}_i(\mathbf r)$ in its vicinity. The particles of fluid component $i$, situated at a position $\mathbf r$ from the centre of the JP, are therefore subjected to an electrochemical force 
	\begin{equation}
	\mathbf F_i(\mathbf r)=-\nabla\tilde{\mu}_i(\mathbf r)=-\nabla\mu_i(\mathbf r)+q_i\mathbf E(\mathbf r),\label{eq:-0}
	\end{equation}
	where $\mu_i(\mathbf r)$ is the chemical potential of fluid component $i$, $q_i$ is the corresponding charge, and $\mathbf E(\mathbf r)$ is the local electric field induced by the chemical activity. The forces $\mathbf F_i(\mathbf r)$ are supposed to be reasonably weak, as to allow for a description of particle motion that is linear in the electrochemical gradients $\nabla\tilde{\mu}_i(\mathbf r)$. The electrochemical force density exerted by the JP on fluid component $i$ is given by
	\begin{equation}
	\bm{\mathcal F}_i(\mathbf r)= n_i(\mathbf r)\mathbf F_i(\mathbf r),\label{eq:-21}
	\end{equation}	
	where $n_i(\mathbf r)$ is the number density of fluid component $i$. Hence, the net electrochemical force density acting on the surrounding fluid can be expressed as
	\begin{equation}
	\bm{\mathcal F}(\mathbf r)=\sum_i \bm{\mathcal F}_i(\mathbf r).\label{eq:-10}
	\end{equation} 
	A specific interaction between fluid component $i$ and the JP surface may further lead to the build-up of an interfacial layer around the JP, whose effective width $\lambda$ is determined by the steepness of the corresponding interaction potential. A schematic representation of such a JP is shown in fig. \ref{fig:-1}.
	
	The condition of LTE has two important consequences for self-diffusiophoretic motion.\cite{Burelbach2018b}	
	First, it implies that the forces acting within the interfacial layer do not induce motion of the JP. Second, it requires that the densities $n_i(\mathbf r)$ in eq. (\ref{eq:-10}) be evaluated to zeroth order in the electrochemical gradients. Hence, we can write
	\begin{equation}
	n_i(\mathbf r)=n_i^b+n_i^\phi(\mathbf r),
	\end{equation}
	where $n_i^\phi(\mathbf r)$ is the interfacial excess density and $n_i^b$ is the constant bulk density of fluid component $i$. As the solvent ($i=0$) is incompressible, we simply have $n_0^\phi(\mathbf r)=0$ and $n_0(\mathbf r)=n_0^b$. For later considerations, it is convenient to introduce the radial distribution function $g_i(\mathbf r)$ of the interfacial excess of solute component $i$ as
	\begin{equation}
	n_i^\phi(\mathbf r)=n_i^bg_i(\mathbf r).\label{eq:-34}
	\end{equation} 
	
	Self-diffusiophoretic motion is an overall force-free transport phenomenon, meaning that it does not lead to a net transport of momentum. As the volume element containing the JP and the fluid is subjected neither to an external force nor to a net hydrodynamic force, the electrochemical forces must obey an action-reaction law of the form \cite{burelbach2019particle}
	\begin{eqnarray}
	\mathbf F +\int \bm{\mathcal F}(\mathbf r) dV=\mathbf 0,\label{eq:-7}
	\end{eqnarray}
	where $\mathbf F$ is the net electrochemical force exerted by the solutes on the JP. An analogous balance equation must hold for the torques induced by the electrochemical gradients, such that
	\begin{eqnarray}
	\bm\tau+\int\mathbf r\times \bm{\mathcal F}(\mathbf r) dV=\mathbf 0,\label{eq:-8}
	\end{eqnarray}
	where $\bm\tau$ is the electrochemical torque exerted by the solutes on the JP.
	
	In order to obtain an Onsager formulation for the active motion of the JP, we consider the average rate of entropy $\sigma_s$ produced by the particle fluxes of all components inside the volume element\cite{burelbach2019particle,DeGroot1963}
	\begin{equation}
	\sigma_s TV=\sum_i\int\mathbf{J}_i(\mathbf r)\cdot\mathbf F_i(\mathbf r)dV+\mathbf{v}\cdot\mathbf F+\bm{\omega}\cdot\bm\tau,\label{eq:-65}
	\end{equation}
	where $\mathbf v$ is the linear and $\bm{\omega}$ is the angular velocity of the JP, $T$ is the temperature of the volume element, and $\mathbf{J}_i(\mathbf r)$ is the local flux of 
	fluid component $i$.
	Based on eq.\ ({\ref{eq:-65}}), Onsager's
	theory of non-equilibrium thermodynamics postulates a linear coupling between the fluxes and electrochemical forces via the phenomenological coefficients $\mathbf{L}_{\alpha \beta}$,\cite{Onsager1931,Onsager1931a} which may have a scalar or tensorial character.\cite{burelbach2019particle}
	For the velocities of the JP, one therefore has
	\begin{equation}
	\mathbf v = \frac{\mathbf F}{\xi_\text{t}}+\sum_i\int\mathbf{L}_{vi}(\mathbf r)\cdot\mathbf F_i(\mathbf r)dV\label{eq:-56}
	\end{equation}
	and
	\begin{equation}
	\bm \omega = \frac{\bm\tau}{\xi_\text{r}}+\sum_i\int\mathbf{L}_{\omega i}(\mathbf r)\cdot\mathbf F_i(\mathbf r)dV,\label{eq:-57}
	\end{equation}
	where 
	\begin{equation}
	\xi_\text{t}=6\pi\eta R \ \ \text{and} \ \ \xi_\text{r}=8\pi\eta R^3\label{eq:-11}
	\end{equation}
	are the translational and rotational friction coefficient of the JP. Similarly, the particle flux of fluid component $i$ 
	must be of the form
	\begin{equation}
	\mathbf{J}_{i}(\mathbf r)=\sum_k\mathbf L_{ik}(\mathbf r)\cdot\mathbf F_k(\mathbf r)+\mathbf L_{iv}(\mathbf r)\cdot\mathbf F+\mathbf L_{i\omega}(\mathbf r)\cdot\bm\tau,\label{eq:-39}
	\end{equation}
	where the index $k$ runs over all fluid components.
	
	We base our approach on the Onsager-Casimir reciprocal relations, which imply that the coupling coefficients $\mathbf{L}_{\alpha \beta}$ are symmetric for linear motion and antisymmetric for angular motion of the JP \cite{Casimir1945,Gaspard2018}
	\begin{equation}
	\mathbf{L}_{iv}(\mathbf r) = \mathbf{L}_{vi}(\mathbf r) \ \ \text{and} \ \ \mathbf{L}_{i\omega}(\mathbf r) = -\mathbf{L}_{\omega i}(\mathbf r).\label{eq:-4}
	\end{equation}
	
	In order to determine $\mathbf{L}_{vi}(\mathbf r)$ and $\mathbf{L}_{\omega i}(\mathbf r)$ from these reciprocal relations, 
	we require a hydrodynamic form for the flux of fluid component $i$ caused by a force and torque on the JP. If $\mathbf u(\mathbf r)$ is the local fluid flow velocity induced by these forces, then the corresponding fluid particle flux $\mathbf J_{i,\mathbf u}(\mathbf r)$ 
	can be written as
	\begin{equation}
	\mathbf J_{i,\mathbf u}(\mathbf r)=n_i(\mathbf r)\mathbf u(\mathbf r).\label{eq:-68}
	\end{equation}
	Within low-Reynolds number hydrodynamics,
	the fluid flow velocity $\mathbf u(\mathbf r)$ is linear in the force $\mathbf F$ and torque $\bm \tau$, such that
	\begin{equation}
	\mathbf u(\mathbf r)=\frac{1}{\xi_\text{t}}\mathbf S(\mathbf r)\cdot\mathbf F+\frac{1}{\xi_\text{r}}\mathbf R(\mathbf r)\cdot\bm \tau,\label{eq:-20}
	\end{equation}
	where $\mathbf S(\mathbf r)$ and $\mathbf R(\mathbf r)$ are the corresponding fluid flow tensors. Using eq. (\ref{eq:-20}) in eq. (\ref{eq:-68}), a comparison with eq. (\ref{eq:-39}) yields
	\begin{equation}
	\mathbf L_{iv}(\mathbf r) = \frac{1}{\xi_\text{t}}n_i(\mathbf r)\mathbf S(\mathbf r) \ \ \text{and} \ \ \mathbf L_{i\omega}(\mathbf r) = \frac{1}{\xi_\text{r}}n_i(\mathbf r)\mathbf R(\mathbf r),\label{eq:-41}
	\end{equation}
	which also determines $\mathbf L_{vi}(\mathbf r)$ and 
	$\mathbf L_{i\omega}(\mathbf r)$ based on the reciprocal relations (\ref{eq:-4}).
	
	For a spherical JP with a non-slip hydrodynamic boundary, the flow tensors have well-known analytical expressions, respectively given by \cite{Landau1987}
	\begin{equation}
	\mathbf S(\mathbf r)=
	\begin{cases}
	\frac{3}{4}\frac{R}{r}\left[\mathbf 1+\hat {\mathbf r}\hat {\mathbf r}-\frac{1}{3}\frac{R^2}{r^2}\left(3\hat {\mathbf r}\hat {\mathbf r}-\mathbf 1\right)\right], &r\geqslant R \\
	\\
	\mathbf 1, &r<R,\label{eq:-44}
	\end{cases}
	\end{equation}
	and
	\begin{equation}
	\mathbf R(\mathbf r)=
	\begin{cases}
	-\frac{R^3}{r^3}\mathbf r\times, &r\geqslant R \\
	\\
	-\mathbf r\times, &r<R,\label{eq:-42}
	\end{cases}
	\end{equation}
	where $r=\left|\mathbf r\right|$ and $\hat {\mathbf r}=\mathbf r/r$. Using eqs. (\ref{eq:-7}), (\ref{eq:-8}) and (\ref{eq:-4}) in eqs. (\ref{eq:-56}) and (\ref{eq:-57}), the linear and angular velocity of the JP can now be expressed as linear functionals of $\bm{\mathcal F}(\mathbf r)$, giving
	\begin{eqnarray}
	\mathbf v\left[\bm{\mathcal F}(\mathbf r) \right]  & = & \frac{1}{\xi_\text{t}}\int_R^\infty\left(\mathbf S(\mathbf r)-\mathbf 1\right)\cdot\bm{\mathcal F}(\mathbf r)dV,\label{eq:-12}\\
	\nonumber\\
	\bm \omega\left[\bm{\mathcal F}(\mathbf r) \right]  & = & -\frac{1}{\xi_\text{r}}\int_R^\infty\left(1-\frac{R^3}{r^3}\right)\mathbf r\times\bm{\mathcal F}(\mathbf r)dV,\label{eq:-23}
	\end{eqnarray}
	where we have directly substituted the expression for $\mathbf R(\mathbf r)$ into the second equation. The notation $\int_R^\infty$ indicates 
	that the volume integral is evaluated from the JP surface to a region in the bulk of the system.
	
\begin{figure}
		\centering{}\includegraphics[width=7.5cm,height=4.5cm]{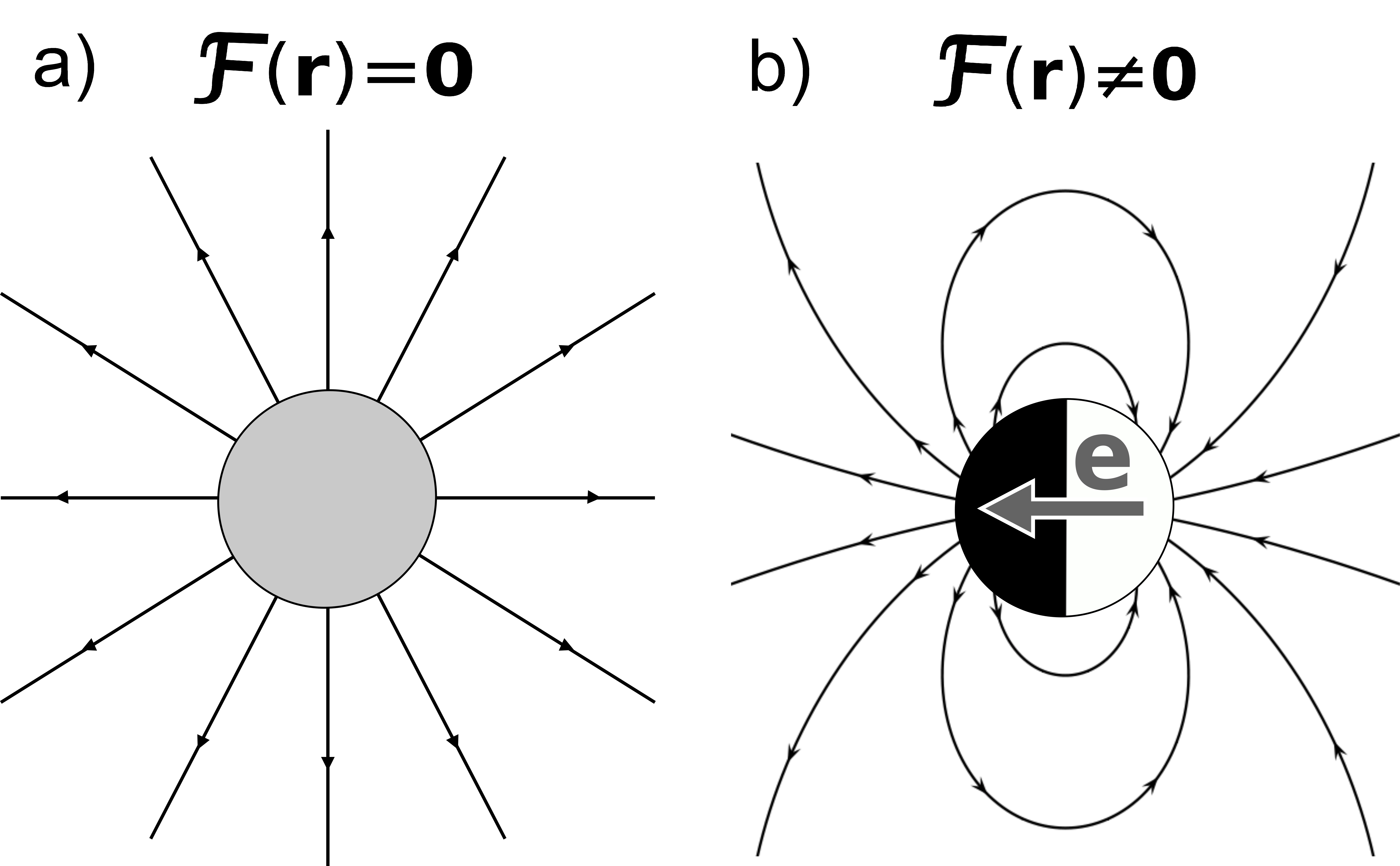}\caption{a) For a spherical JP with an isotropic chemical activity, the electrochemical field of a reacting solute is radially symmetric, as shown by the thin black lines. Due to the no-flux boundary condition at the JP surface, the solvent can induce an opposing radial gradient in its chemical potential, which guarantees a perfect hydrostatic equilibrium around the JP. b) However, this hydrostatic equilibrium is broken if the chemical activity of the JP is anisotropic, as for half-capped JPs. In particular, a half-capped JP is axisymmetric and can therefore be assigned a directional unit vector $\mathbf e$, which is chosen to point towards the chemically active hemisphere (shown in black).}\label{fig:-2}
	\end{figure}

	The solvent does not participate in a chemical reaction and is therefore in principle capable of maintaining a hydrostatic equilibrium around the JP. However, the solvent can only maintain a hydrostatic equilibrium normal to the surface of a spherical JP if the electrochemical forces on the solutes are radially symmetric (see fig. \ref{fig:-2}). It is therefore instructive to write the electrochemical force density as
	$\bm{\mathcal F}(\mathbf r)=\bm{\mathcal F}^\circ(r)+\left(\bm{\mathcal F}(\mathbf r)-\bm{\mathcal F}^\circ(r)\right)$,
	where the radially symmetric component $\bm{\mathcal F}^\circ(r)=\mathcal F^\circ(r)\hat {\mathbf r}$ vanishes if a hydrostatic equilibrium is maintained normal to the surface. The electrochemical force density $\bm{\mathcal F}_0(\mathbf r)=\bm{\mathcal F}_0^\circ(r)$ on the solvent is thus fixed by the condition $\bm{\mathcal F}^\circ(r)=\sum_i\bm{\mathcal F}_i^\circ(r)=\mathbf 0$, such that $\bm{\mathcal F}_0^\circ(r)=-\sum_{i\neq 0}\bm{\mathcal F}_i^\circ(r)$. Using this to eliminate $\bm{\mathcal F}_0(\mathbf r)$ in eq. (\ref{eq:-10}), the net electrochemical force density can be expressed as 
	\begin{equation}
	\bm{\mathcal F}(\mathbf r)=\sum_{i\neq 0}\left[\bm{\mathcal F}_i(\mathbf r)-\bm{\mathcal F}_i^\circ(r) \right].\label{eq:-25}
	\end{equation}
	However, the functional forms given by eqs. (\ref{eq:-12}) and (\ref{eq:-23}) vanish under radial symmetry. As a result, the radially symmetric components in eq. (\ref{eq:-25}) do not contribute to the active motion of the JP, which implies that $\mathbf v\left[\bm{\mathcal F}(\mathbf r) \right]=\sum_{i\neq 0}\mathbf v\left[\bm{\mathcal F}_i(\mathbf r)\right]$ and $\bm\omega\left[\bm{\mathcal F}(\mathbf r) \right]=\sum_{i\neq 0}\bm\omega\left[\bm{\mathcal F}_i(\mathbf r)\right]$. Using this and eq. (\ref{eq:-21}) in eqs. (\ref{eq:-12}) and (\ref{eq:-23}), the active linear and angular velocity $\mathbf v_\text{a}$ and $\bm \omega_\text{a}$ of the JP take the final forms
	\begin{eqnarray}
	\mathbf v_\text{a} & = & \frac{1}{\xi_\text{t}}\sum_{i\neq 0}\int_R^\infty n_i(\mathbf r)\left(\mathbf S(\mathbf r)-\mathbf 1\right)\cdot\mathbf F_i(\mathbf r)dV,\label{eq:-3}\\
	\nonumber\\
	\bm \omega_\text{a} & = & -\frac{1}{\xi_\text{r}}\sum_{i\neq 0}\int_R^\infty n_i(\mathbf r)\left(1-\frac{R^3}{r^3}\right)\mathbf r\times\mathbf F_i(\mathbf r)dV,\nonumber\\
	\label{eq:-14}
	\end{eqnarray}
	which only refer to the electrochemical forces $\mathbf F_i(\mathbf r)$ acting on the solutes ($i\neq 0$). Equations (\ref{eq:-3}) and (\ref{eq:-14}) 
	can now be used to determine the active motion of single JPs when the local solute densities and electrochemical forces 
	are known.
	
	It is instructive to compare eqs. (\ref{eq:-3}) and (\ref{eq:-14}) to the results obtained within the boundary-layer approximation. This approximation relates the velocity of a particle to the interfacial excess densities $n_i^\phi(\mathbf r)$ of the solutes, by assuming that the range of the specific interaction between the JP and the solutes is very short compared to the JP radius ($\lambda\ll R$). This is indeed well-justified for phoretic motion of a passive particle subjected to uniform electrochemical bulk gradients, which couple to the interfacial solute layer to break the local hydrostatic equilibrium at the surface. However, as shown in fig. \ref{fig:-2}, an active JP may also break the hydrostatic equilibrium without interfacial solute excess if its chemical activity is anisotropic.
	
	In order to recover the boundary-layer treatment of active motion, the net solute densities $n_i(\mathbf r)$ in eqs. (\ref{eq:-3}) and (\ref{eq:-14}) must therefore be replaced by the interfacial excess densities $n_i^\phi(\mathbf r)$. 
	A first order expansion in the small parameter $z/R=(r-R)/R\ll 1$
	is then performed, yielding
	$\mathbf S(\mathbf r)-\mathbf 1\approx-\frac{3}{2}\frac{z}{R}\left(\mathbf 1-\hat {\mathbf r}\hat {\mathbf r}\right)$ and $\left(1-R^3/r^3\right)\mathbf r\approx 3z\hat {\mathbf r}$, where $z$ is the radial distance from the JP surface. The volume integral can further be written as $\int_R^\infty(...)dV\approx4\pi R^2\left\langle\int_0^\infty(...)dz\right\rangle_S$, where $\left\langle...\right\rangle_S\equiv\frac{1}{4\pi}\oint_S \left(...\right) \sin\theta d\theta d\varphi$ is the average over the surface $S$ of the JP. As electrochemical forces can be assumed independent of $z$ inside thin interfacial layers, we further have $\mathbf F_i(\mathbf r)\approx\mathbf F_i(\hat {\mathbf r})$.
	With this and eqs. (\ref{eq:-34}) and (\ref{eq:-11}), eqs. (\ref{eq:-3}) and (\ref{eq:-14}) finally reduce to
	\begin{eqnarray}
	\mathbf v_\text{a} & = & -\sum_{i\neq 0}\left\langle M_i(\hat {\mathbf r})\left(\mathbf 1-\hat {\mathbf r}\hat {\mathbf r}\right)\cdot n_i^b\mathbf F_i(\hat {\mathbf r})\right\rangle_S,\label{eq:-15}\\
	\nonumber\\
	\bm \omega_\text{a} & = & -\frac{3}{2 R}\sum_{i\neq 0}\left\langle M_i(\hat {\mathbf r})\hat {\mathbf r}\times n_i^b\mathbf F_i(\hat {\mathbf r})\right\rangle_S,\label{eq:-16}
	\end{eqnarray}
	where
	\begin{equation}
	M_i(\hat {\mathbf r}) = \frac{1}{\eta}\int_0^\infty g_i(\mathbf r)zdz\label{eq:-61}
	\end{equation}
    can be defined as the local phoretic surface mobility of the JP due to its specific interaction with solute component $i$.
	
    The form on the RHS of eq. (\ref{eq:-15}) coincides with the formal definition of the fluid slip velocity $\mathbf u_{\text{slip}}(\hat {\mathbf r})$, which is commonly used for a description of phoretic motion within the boundary-layer approximation \cite{Anderson1989,Derjaguin1987}
	\begin{equation}
	\mathbf u_{\text{slip}}(\hat {\mathbf r}) = \sum_{i\neq 0}M_i(\hat {\mathbf r})\left(\mathbf 1-\hat {\mathbf r}\hat {\mathbf r}\right)\cdot n_i^b\mathbf F_i(\hat {\mathbf r}).\label{eq:-59}
	\end{equation}
    Noting that for any vector field $\mathbf A(\mathbf r)$ we have $\hat {\mathbf r}\times\mathbf A(\mathbf r)=\hat {\mathbf r}\times\left[\left(\mathbf 1-\hat {\mathbf r}\hat {\mathbf r}\right)\cdot\mathbf A(\mathbf r)\right]$, we can use eq. (\ref{eq:-59}) to rewrite eqs. (\ref{eq:-15}) and (\ref{eq:-16}) as
	\begin{eqnarray}
	\mathbf v_\text{a} & = & -\left\langle\mathbf u_{\text{slip}}(\hat {\mathbf r})\right\rangle_S\label{eq:-1}\\
	\nonumber\\
	\bm \omega_\text{a} & = & -\frac{3}{2R}\left\langle\hat {\mathbf r}\times \mathbf u_{\text{slip}}(\hat {\mathbf r})\right\rangle_S.\label{eq:-2}
	\end{eqnarray}

	With eqs. (\ref{eq:-1}) and (\ref{eq:-2}), we have recovered the standard forms of the linear and angular velocities as previously obtained within the boundary-layer approximation.\cite{Anderson1989,Bickel2014}
	
	\subsection{Diffusiophoretic interactions between Janus particles}
	
	The phase behaviour of an active suspension is determined by the relative motion of the JPs, induced by diffusiophoretic interactions between them. The reciprocal approach presented in section \ref{sec:-1} can directly be applied to this relative motion if hydrodynamic interactions and mutual boundary conditions are ignored. The latter condition implies that the electrochemical fields created by one JP do not have to satisfy any boundary conditions at the surface of another JP. Here, we will therefore provide a reciprocal description of diffusiophoretic interactions that holds in the far-field regime of an active suspension, when the separations between the JPs are reasonably large compared to effective diameter $2(R+\lambda)$. As a JP is well approximated by a chemical monopole in the far-field, the fluid can be assumed at hydrostatic equilibrium far away from its surface. At large distances from a JP, where $n_i(\mathbf r)=n_i^b$, the electrochemical force $\bm{\mathcal F}_0(\mathbf r)$ on the solvent is therefore fixed by the condition
	\begin{equation}
	\bm{\mathcal F}_0(\mathbf r)=-\sum_{i\neq 0}n_i^b\mathbf F_i^\circ(r),\label{eq:-26}
	\end{equation}
	where $\mathbf F_i^\circ(r)$ are the radially symmetric electrochemical forces exerted by the JP on the solutes far away from its surface. 
	
	\begin{figure}
		\centering{}\includegraphics[width=4.8cm,height=4cm]{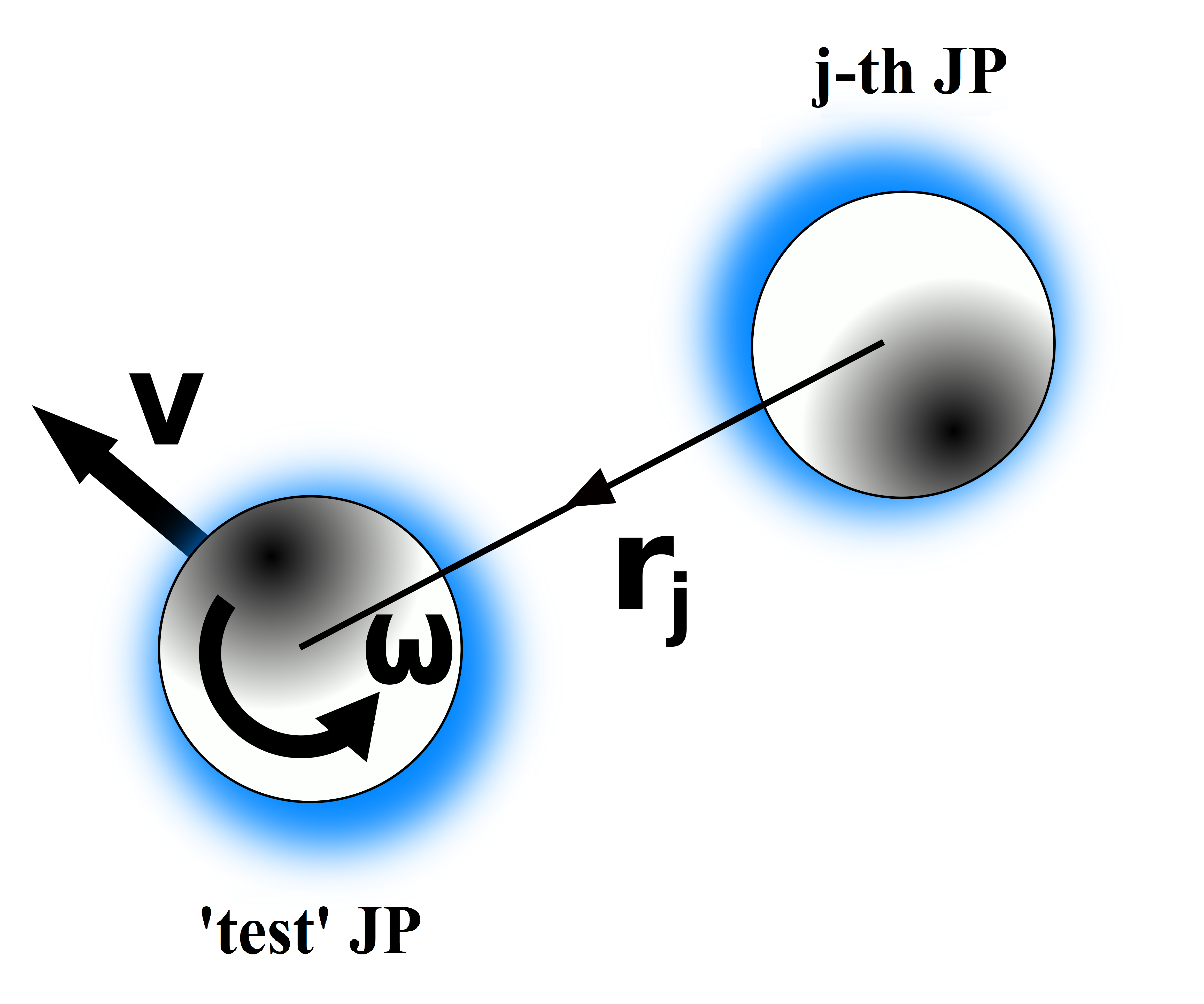}\caption{The net velocities $\mathbf v$ and $\bm\omega$ of a 'test' JP differ from its active velocities $\mathbf v_\text{a}$ and $\bm\omega_\text{a}$ if it diffusiophoretically interacts with another ($j$-th) JP.}\label{fig:-3}
	\end{figure}
	
	Let us now consider an active suspension of $N$ JPs indexed by the letter $j$, at positions $\bm{\mathcal R}_j$ inside the system. Within the far-field approximation, we denote the radially symmetric electrochemical force exerted by the $j$-th JP on solute $i$ at a position $\bm{\mathcal R}$ far away from its surface by $\mathbf F_{ij}^\circ(r_j)$, where $r_j=\left| \bm{\mathcal R}-\bm{\mathcal R}_j\right|\gg 2(R+\lambda)$.
	The net electrochemical force $\mathbf F_i^N(\bm{\mathcal R})$ exerted by all $N$ JPs on solute $i$ at position $\bm{\mathcal R}$ is therefore given by
	\begin{equation}
	\mathbf F_i^N(\bm{\mathcal R})=\sum_j\mathbf F_{ij}^\circ(r_j).\label{eq:-50}
	\end{equation} 
	In view of eq. (\ref{eq:-0}), this net electrochemical force can also be written as
	\begin{equation}
	\mathbf F_i^N(\bm{\mathcal R})=-\nabla\mu_i(\bm{\mathcal R})+q_i\mathbf E(\bm{\mathcal R}),\label{eq:-60}
	\end{equation}
	where $\mathbf E(\bm{\mathcal R})$ and $\mu_i(\bm{\mathcal R})$ are to be interpreted as the net electric field and chemical potential of solute $i$ induced by the $N$ JPs at position $\bm{\mathcal R}$.
	Based on eq. (\ref{eq:-26}), the electrochemical force on the solvent at position $\bm{\mathcal R}$ is hence fixed by
	\begin{equation}
	\bm{\mathcal F}_0(\bm{\mathcal R})=-\sum_{i\neq 0}n_i^b\mathbf F_i^N(\bm{\mathcal R}).\label{eq:-33}
	\end{equation}
	If another 'test' JP is now placed at position $\bm{\mathcal R}$ (as shown in fig. \ref{fig:-3}), then the specific interaction between this JP and solute component $i$ changes the density of that solute from $n_i^b$ to $n_i(\mathbf r)$ at a position $\mathbf r$ from its centre, thus inducing a corresponding force density 
	$n_i(\mathbf r)\mathbf F_i^N(\bm{\mathcal R})$.
	With eq. (\ref{eq:-33}), the net electrochemical force density resulting from the coupling of the forces $\mathbf F_i^N(\bm{\mathcal R})$ to the local solute densities $n_i(\mathbf r)$ around the test JP can therefore be expressed as
	\begin{eqnarray}
	\bm{\mathcal F}_d(\mathbf r)&=&\bm{\mathcal F}_0(\bm{\mathcal R})+\sum_{i\neq 0}n_i(\mathbf r)\mathbf F_i^N(\bm{\mathcal R})\nonumber\\
	&=&\sum_{i\neq 0}n_i^\phi(\mathbf r)\mathbf F_i^N(\bm{\mathcal R}),
	\end{eqnarray}
    where $n_i^\phi(\mathbf r)=n_i(\mathbf r)-n_i^b$ is the interfacial excess density of solute $i$ at the surface of the test JP.
	
	As the test JP and the surrounding fluid at position $\bm{\mathcal R}$ are not subjected to a net external or hydrodynamic force, the force density $\bm{\mathcal F}_d(\mathbf r)$ satisfies the same action-reaction laws as given by eqs. (\ref{eq:-7}) and (\ref{eq:-8}). Applying the reciprocal approach from section \ref{sec:-1}, the linear and angular velocity $\mathbf v_d$ and $\bm \omega_d$ of the test JP induced by its diffusiophoretic interaction with the other JPs take the forms
	\begin{eqnarray}
	\mathbf v_d & = & \frac{1}{\xi_\text{t}}\sum_{i\neq 0}\int_R^\infty n_i^\phi(\mathbf r)\left(\mathbf S(\mathbf r)-\mathbf 1\right)dV\cdot\mathbf F_i^N(\bm{\mathcal R}),\label{eq:-45}\\
	\nonumber\\
	\bm \omega_d & = & -\frac{1}{\xi_\text{r}}\sum_{i\neq 0}\int_R^\infty n_i^\phi(\mathbf r)\left(1-\frac{R^3}{r^3}\right)\mathbf rdV\times\mathbf F_i^N(\bm{\mathcal R}).\nonumber\\
	\label{eq:-46}
	\end{eqnarray}
    From a comparison of eqs. (\ref{eq:-45}) and (\ref{eq:-46}) to eqs. (\ref{eq:-3}) and (\ref{eq:-14}), it becomes clear that 
	motion induced by far-field 
	diffusiophoretic interactions distinguishes itself from active motion of single JPs in that it relies on an interfacial solute excess at the JP surface, similar to phoretic motion of passive particles.

	\section{Model calculation: Half-capped Janus particles}
	
	For our model calculations, we consider JPs that catalyse a chemical reaction of solutes at their surface.\cite{Golestanian2005,Gaspard2018}	
	The electrochemical fields created by a JP are determined by the continuity equations of the solutes. The stationary forms of these equations read
	\begin{equation}
	\nabla\cdot \mathbf J_i(\mathbf r)=\sigma_i(\mathbf r),\label{eq:-5}
	\end{equation}
	where $\sigma_i(\mathbf r)$ is the corresponding chemical source density distribution located on the surface of
	the JP. To describe the chemical reaction, each solute component is assigned a stoichiometric coefficient $\nu_i$, which is negative for reactants and positive for products. If a product with a stoichiometric coefficient $\nu_i=1$ is produced at a local rate $\sigma(\mathbf r)$ per unit volume, then the reaction satisfies
	\begin{equation}
	\sigma_i(\mathbf r)=\nu_i\sigma(\mathbf r).\label{eq:-55}
	\end{equation}
	To solve eq. (\ref{eq:-5}), we assume that the motion of the solutes is diffusion-dominated. This implies that the cross-coefficients in eq. (\ref{eq:-39}) are negligible compared to the diagonal coefficient $\mathbf L_{ii}(\mathbf r)$, which is described by the scalar relation
    \begin{equation}
    \mathbf L_{ii}(\mathbf r)=L_{ii}(\mathbf r)=\frac{n_i(\mathbf r)}{\xi_i},
    \end{equation}
	where
	\begin{equation}
	\xi_i=6\pi\eta R_i \label{eq:-13}
	\end{equation}
	is the translational friction coefficient of a particle of fluid component $i$, with a hydrodynamic radius $R_i$. Using $ \mathbf J_i(\mathbf r)\approx n_i(\mathbf r)\mathbf F_i(\mathbf r)/\xi_i$ and eq. (\ref{eq:-55}) in eq. (\ref{eq:-5}), we obtain
	\begin{equation}
	\nabla\cdot\frac{n_i(\mathbf r)}{\xi_{i}}\mathbf F_i(\mathbf r)=\nu_i\sigma(\mathbf r).\label{eq:-6}
	\end{equation}
	As the solute densities are evaluated to zeroth order in the electrochemical gradients, a position dependence of $n_i(\mathbf r)$ exclusively stems from the specific interaction between the solutes and the JP surface. Here, we treat the electrochemical forces $\mathbf F_i(\mathbf r)$ as decoupled from the interfacial solute layers, by requiring that the interfacial excess density of a solute is weak compared to its bulk density: $\left| n_i^\phi(\mathbf r)\right| \ll n_i^b$. In this case, the factor $n_i(\mathbf r)/\xi_{i}$ can be assumed constant such that $n_i(\mathbf r)/\xi_{i}\approx n_i^b/\xi_{i}$. Equation (\ref{eq:-6}) then reduces to a Poisson equation
	for the electrochemical fields $\tilde{\mu}_i(\mathbf r)$ introduced in eq. (\ref{eq:-0}), which can be solved by a multipole expansion.
	
	\subsection{Active motion of a single half-capped Janus particle}
	
	A system commonly studied theoretically and experimentally is that of half-capped JPs. As shown in fig. \ref{fig:-2}b, a half-capped JP has an upper hemisphere ($+$) with chemically active cap, and a passive lower hemisphere ($-$). Due to its axisymmetry, it can further be assigned a directional unit vector $\mathbf e$, which is chosen to point from the passive to the active hemisphere. The reaction exclusively occurs on the surface of the cap with a constant production rate $\sigma$ per unit area. Assuming that the solutes are much smaller than the JP ($R_i\ll R$) and that the reaction rate is limited by the number of catalytic sites on the cap,\cite{Golestanian2007} the chemical source density distribution is simply given by
	\begin{equation}
	\sigma_i(\mathbf r)=
	\begin{cases}
	\nu_i\sigma\delta(r-R), &0\leqslant\theta\leqslant \frac{\pi}{2} \\
	\\
	0, &\frac{\pi}{2}<\theta\leqslant\pi,\label{eq:-47}
	\end{cases}
	\end{equation}
	where $\cos\theta=\mathbf e\cdot\mathbf r$ and $\sigma>0$. 
	Moreover, the hemispheres may specifically interact with the solutes via different interactions potentials. Here, we assume 
	that these potentials undergo a sharp transition at the equatorial plane of the JP ($\theta=\pi/2$). The interfacial excess 
	densities of the solutes can then approximately
	be described by different radial distribution functions $g_i^\pm(r)$ on each side, such that
	\begin{equation}
	n_i^\phi(\mathbf r)=
	\begin{cases}
	n_i^bg_i^+(r), &0\leqslant\theta\leqslant \frac{\pi}{2} \\
	\\
	n_i^bg_i^-(r), &\frac{\pi}{2}<\theta\leqslant\pi.\label{eq:-24}
	\end{cases}
	\end{equation}
	To describe the motion of half-capped JPs, it also turns out instructive to introduce the functions
	\begin{equation}
	\bar{g}_i(r)=\frac{1}{2}\left[g_i^+(r)+g_i^-(r)\right]
	\end{equation}
	and
	\begin{equation}
	\delta g_i(r)=\frac{1}{2}\left[g_i^+(r)-g_i^-(r)\right],
	\end{equation}
	which respectively quantify the radial symmetry and symmetry-breaking of the interfacial solute layer.
	
	As already mentioned, the electrochemical forces $\mathbf F_i(\mathbf r)$ are determined from eq. (\ref{eq:-6}) by 
	requiring that $\left|g_i^\pm(r)\right|\ll 1$. The resulting Poisson equation has previously been solved for self-thermophoretic 
	JPs with a source distribution given by eq. (\ref{eq:-47}).\cite{Bickel2013} For self-diffusiophoretic JPs, the corresponding electrochemical forces can be written as 
	\begin{equation}
	\mathbf F_i(\mathbf r)=\frac{1}{2}\frac{\xi_i}{n_i^b}\nu_i\sigma\bm f(\mathbf r),\label{eq:-27}
	\end{equation}
	where the rescaled force $\bm f(\mathbf r)$ is a polynomial expansion of the form
	\begin{equation}
	\begin{split}
	&\bm f(\mathbf r)=\\
	&\sum_{m=0}^\infty\alpha_{m,i}\left(\frac{R}{r}\right)^{m+2}\left[(m+1)P_m(c_\theta)\hat{\mathbf r}+s_\theta P'_m(c_\theta)\hat{\bm\theta}\right].\label{eq:-28}
	\end{split}
	\end{equation}
	Here, $P_m(x)$ is the Legendre polynomial of degree $m$ and $P'_m(x)=\partial P_m(x)/\partial x$. We have further used the short-hand notation $c_\theta\equiv\cos\theta$ and $s_\theta\equiv\sin\theta$.
	Assuming that the solutes cannot penetrate the JP surface, the coefficients $\alpha_{m,i}$ are given by\cite{Bickel2013}
	\begin{equation}
	\alpha_{2l,i} = \delta_{2l,0}
	\end{equation}
	if $m=2l$ is even ($l\in\mathbb N_0$) and 
	\begin{equation}
	\alpha_{2l+1,i} = \frac{(-1)^l(2l)!(4l+3)}{2^{2l+2}\left[(l+1)!\right]^2},
	\end{equation}
	if $m=2l+1$ is odd. 
	
	Due to their axisymmetry, it is clear that half-capped JPs cannot undergo active rotation, hence 
	\begin{equation}
	\bm{\omega}_a=\mathbf 0.
	\end{equation} 
	As $\hat{\mathbf r}$ and $\hat{\bm \theta}$ are eigenvectors of $\mathbf S(\mathbf r)-\mathbf 1$, the evaluation of eq.\ (\ref{eq:-3}) for the active linear velocity $\mathbf v_\text{a}$ involves surface averages over the vectors $P_m(c_\theta)\hat{\mathbf r}$ and $s_\theta P'_m(c_\theta)\hat{\bm\theta}$. Based on the orthogonality of the Legendre polynomials, these surface averages are found to have a non-zero contribution from the chemical dipole ($m=1$) only. Evaluating eq. (\ref{eq:-3}) using eqs. (\ref{eq:-44}), (\ref{eq:-24}), (\ref{eq:-27}) and (\ref{eq:-28})
	together with $n_i(\mathbf r) = n_i^b + n_i^\phi(\mathbf r)$,	
	the active linear velocity of a half-capped JP can finally be expressed as 
	\begin{equation}
	\mathbf v_\text{a}=(v_0+v_\phi)\mathbf e,\label{eq:-58}
	\end{equation}
	where
	\begin{eqnarray}
	v_0&=&\frac{1}{2}\pi R^2\sigma\sum_{i\neq 0}\nu_iR_i,\label{eq:-48}\\
	\nonumber\\
	v_\phi&=&\frac{3}{4}\pi R\sigma\sum_{i\neq 0}\nu_iR_i\int_R^\infty\frac{R^2}{r^2}\left(1-\frac{R^2}{r^2}\right)\bar{g}_i(r)dr.\nonumber\\
	\label{eq:-49}
	\end{eqnarray}
	
	Based on eqs. (\ref{eq:-61}) and (\ref{eq:-24}), the phoretic surface mobility on each hemisphere can be written as 
	\begin{equation}
	M_i^\pm = \frac{1}{\eta}\int_0^\infty g_i^\pm(z)zdz.\label{eq:-32}
	\end{equation}
	As a result, a first order expansion in $z/R\ll 1$ of eq. (\ref{eq:-49}) yields
    \begin{equation}
    v_\phi=\frac{3}{4}\pi\eta\sigma\sum_{i\neq 0}\nu_iR_i\left( M_i^++M_i^-\right),\label{eq:-62}
    \end{equation}
    which is in agreement with the result obtained from a previous boundary-layer treatment for half-capped JPs.\cite{Golestanian2007} This becomes evident by noting that $n_i^b\mathbf F_i(\mathbf r)=-k_BT\nabla n_i(\mathbf r)$ if a solute component $i$ behaves like an ideal gas, where $k_B$ is the Boltzmann constant. Using the substitutions $D=k_BT/(6\pi\eta R_i)$, $\alpha_+=\nu_i\sigma$, $\alpha_-=0$ and $\mu_\pm=-M_i^\pm k_BT$ in eq. (9) of this work then allows the recovery of eq. (\ref{eq:-62}).

	Two important conclusions can be drawn from eqs.\ (\ref{eq:-48}) and (\ref{eq:-49}). Unlike phoretic motion of chemically passive particles,\cite{Anderson1985} eq. (\ref{eq:-48}) suggests that active motion can occur in the absence of an interfacial solute layer around a JP if the hydrostatic equilibrium at its surface is broken by an anisotropic chemical activity. For a chemical reaction described by eq. (\ref{eq:-47}), the corresponding coefficient $v_0$ is proportional to the stoichiometrically weighted sum of the hydrodynamic solute radii ($\sum_{i\neq 0}\nu_iR_i$). However, the coefficient $v_\phi$ given by eq. (\ref{eq:-49}) relies on an interfacial solute layer and vanishes if $\bar{g}_i(r)=0$. For purely electrostatic interactions, the latter conclusion implies that no active motion is induced by the coupling of electrochemical forces to an interfacial solute layer if the hemispheres of the JP have an equal and opposite charge distribution.
	
	A direct comparison of eqs. (\ref{eq:-58}), (\ref{eq:-48}) and (\ref{eq:-49}) to experiments is challenging as it requires knowledge of the reaction rate and the radial distributions functions $g_i^\pm(r)$, which are often not precisely known for all the solutes. However, a particularly simple case occurs if a single solute component $A$ with a hydrodynamic radius $R_A$ undergoes a conformational change at the JP surface, yielding a product $B$ with a different hydrodynamic radius $R_B$. If the specific interactions of solutes $A$ and $B$ with the JP surface are weak, then the interfacial contribution can be neglected ($v_\phi\approx 0$) and the active velocity of the JP reduces to
	\begin{equation}
	\mathbf v_\text{a}=v_0\mathbf e=\frac{1}{2}\pi R^2\sigma\left(R_B-R_A\right)\mathbf e.
	\end{equation}
	The order of magnitude of the production rate per unit area $\sigma$ depends on the considered reaction and the catalytic properties of the active cap. For a micron-sized JP ($R\sim\SI{1}{\micro\meter}$) whose cap changes the hydrodynamic solute radius by $\left| R_B-R_A\right|\sim\SI{0.1}{\angstrom}$, a production rate per unit area of just $\sigma\sim 10^{3}$ \SI{}{\second}$^{-1}$\SI{}{\micro\meter}$^{-2}$ would yield active velocities in the experimental range of several \SI{}{\micro\meter} \SI{}{\second}$^{-1}$.
	
	\subsection{Motion induced by diffusiophoretic interactions between half-capped Janus particles}		
	
	We now address the velocity of a half-capped JP induced by diffusiophoretic interactions with other JPs. To this end, we introduce the net rescaled force $\bm f_N(\bm{\mathcal R})$ exerted by $N$ surrounding JPs on the solutes in the vicinity of another JP at position $\bm{\mathcal R}$ via
	\begin{equation}
	\mathbf F_i^N(\bm{\mathcal R})=\frac{1}{2}\frac{\xi_i}{n_i^b}\nu_i\sigma\bm f_N(\bm{\mathcal R}).\label{eq:-29}
	\end{equation}
	Within the far-field approximation upon which eqs. (\ref{eq:-45}) and (\ref{eq:-46}) are based, only the contribution $\bm f^\circ(r)=\left(R/r\right)^{2}\hat{\mathbf r}$ from the chemical monopole ($m=0$) should be kept in eq. (\ref{eq:-28}), such that
	\begin{equation}
	\bm f_N(\bm{\mathcal R})=\sum_j \bm f^\circ(r_j)=\sum_j\left(\frac{R}{r_j}\right)^{2}\hat{\mathbf r}_j,\label{eq:-30}
	\end{equation}
	where $r_j=\left|\bm{\mathcal R}-\bm{\mathcal R}_j \right|$ and $\hat{\mathbf r}_j=(\bm{\mathcal R}-\bm{\mathcal R}_j)/r_j$. Diffusiophoretic interactions are thus expected to dominate over hydrodynamic interactions in the far-field regime, when fluid flows induced by force-free motion decay with distance as $1/r^3$.\cite{Anderson1989,Yang2014} Even in the presence of a hydrodynamic force-dipole contribution, which decays as $1/r^2$, ref.\cite{Liebchen2019} argues that in realistic systems diffusiophoretic interactions should be more important.
	
	Evaluating eqs. (\ref{eq:-45}) and (\ref{eq:-46}) with eq. (\ref{eq:-29}) yields
	\begin{equation}
	\mathbf v_d=v_d\bm f_N(\bm{\mathcal R}) \ \ \text{and} \ \ \bm{\omega}_d=\omega_d\mathbf e\times \bm f_N(\bm{\mathcal R}),
	\end{equation}
	where
	\begin{eqnarray}
	v_d & = & -\ 2\pi\sigma\sum_{i\neq 0}\nu_iR_i\int_R^\infty\left(\frac{r^2}{R}-r\right)
	\bar{g}_i(r)dr,\label{eq:-51}\\
	\nonumber\\
	\omega_d & = & -\frac{3}{4}\pi\sigma\sum_{i\neq 0}\nu_iR_i\int_R^\infty\left(\frac{r^3}{R^3}-1\right)\delta g_i(r)dr.\label{eq:-52}
	\end{eqnarray}
	Although it can be seen from eqs. (\ref{eq:-48}), (\ref{eq:-49}), (\ref{eq:-51}) and (\ref{eq:-52}) that the coefficients $v_0$, $v_\phi$, $v_d$ and $\omega_d$ depend on similar chemical, interfacial and hydrodynamic properties, it is in general not possible to obtain a direct relation between them. The coefficients $v_\phi$ and $v_d$ are both related to the radial function $\bar{g}_i(r)$, but the velocities $\mathbf v_\text{a}$ and $\mathbf v_d$ can nonetheless be tuned independently due to the additional coefficient $v_0$ in $\mathbf v_\text{a}$. Moreover, a free tuning of the angular velocity $\bm{\omega}_d$ is possible due to its dependence on $\delta g_i(r)$ rather than $\bar{g}_i(r)$.
	
	As previously shown,\cite{Liebchen2017} however, the coefficients $v_d$ and $\omega_d$ can be brought into direct relation with the interfacial coefficient $v_\phi$ of the active linear velocity within the boundary-layer approximation. Performing a first order expansion in $z/R\ll 1$ and using eq. (\ref{eq:-32}), eqs. (\ref{eq:-51}) and (\ref{eq:-52}) simplify to
	\begin{equation}
	v_d=-\pi\eta\sigma\sum_{i\neq 0}\nu_iR_i\left(M_i^++M_i^- \right)
	\end{equation}
	and
	\begin{equation}
	\omega_d=-\pi\eta\sigma\frac{9}{8R}\sum_{i\neq 0}\nu_iR_i\left(M_i^+-M_i^-\right).
	\end{equation}
	Provided that only one of the hemispheres specifically interacts with the solutes ($M_i^+=0$ or $M_i^-=0$), we obtain
	\begin{equation}
	v_d=-\frac{4}{3}v_\phi \ \ \text{and} \ \ \omega_d=\mp\frac{3}{2}\frac{v_\phi}{R},\label{eq:-53}
	\end{equation}
	where the interfacial coefficient $v_\phi$ is given by eq. (\ref{eq:-62}). Here, the minus sign in $\mp$ for $\omega_d$ applies if $M_i^-=0$ and the plus sign applies if $M_i^+=0$.
	The net linear and angular velocity $\mathbf v=\mathbf v_\text{a}+\mathbf v_d$ and $\bm{\omega}=\bm{\omega}_d$ take particularly simple forms if we further require that $v_\phi\gg v_0$. Using eq. (\ref{eq:-53}) and $\mathbf v_\text{a}=v_\phi\mathbf e$, we then obtain
	\begin{equation}
	\mathbf v=v_\phi\left( \mathbf e -\frac{4}{3}\bm f_N(\bm{\mathcal R})\right)  \ \ \text{and} \ \ \bm{\omega}=\mp\frac{3v_\phi}{2R}\mathbf e\times \bm f_N(\bm{\mathcal R}).\label{eq:-54}
	\end{equation}
	In the far-field, one has
	$\left|\bm f_N(\bm{\mathcal R})\right|\ll 1$, meaning that the net linear velocity $\mathbf v$ nearly coincides with the active velocity $\mathbf v_\text{a}=v_\phi\mathbf e$. Under this assumption, eq. (\ref{eq:-54}) can be expressed as
	\begin{equation}
	\mathbf v=v_\phi\mathbf e  \ \ \text{and} \ \ \bm{\omega}=\pm\frac{3}{2R}\sum_j\left(\frac{R}{r_j}\right)^{2}\hat{\mathbf r}_j\times \mathbf v,\label{eq:-31}
	\end{equation}
	where we have also substituted eq. (\ref{eq:-30}). 
	
	The rotational behaviour described by eq. (\ref{eq:-31}) agrees with previous observations.\cite{Liebchen2017} If $\pm$ is positive, then the test JP has an interfacial solute excess on the capped hemisphere ($M_i^-=0$) and tends to rotate its linear velocity $\mathbf v$ towards another ($j$-th) JP. In this case, the diffusiophoretic interaction between two JPs is termed 'chemoattractive'. If $\pm$ is negative, then the test JP has an interfacial solute excess on the passive hemisphere ($M_i^+=0$) and tends to rotate away from another ($j$-th) JP. The diffusiophoretic interaction is then said to be 'chemorepulsive'. 
	
	\section{An example of active rotation: Charged Janus particle with a non-uniform zeta potential}
	
	To evidence the possibility
	of active rotation, we consider a weakly charged JP with an anisotropic surface charge distribution. For an axisymmetric chemical source density $\sigma_i(\mathbf r)$, a first-order multipole expansion of the electrochemical fields in eq. (\ref{eq:-6}) yields the following form for the corresponding forces 
	\begin{equation}
	\mathbf F_i(\mathbf r)=\frac{\xi_{i}}{4\pi n_i^b}\left[\frac{k_i\hat{\mathbf r}}{r^2}+\frac{1}{r^3}\left(3\hat {\mathbf r}\hat {\mathbf r}-\mathbf 1\right)\cdot\mathbf p_i\right].\label{eq:-9}
	\end{equation}
	The chemical dipole moment $\mathbf p_i$ quantifies a weak anisotropy in the chemical activity of the JP surface, which consumes/produces particles of solute $i$ at an average rate $k_i$. In view of eq. (\ref{eq:-55}), the reaction satisfies $k_i=\nu_i k$ and $ \mathbf p_i=\nu_i\mathbf p$, where $k$ and $\mathbf p$ are the corresponding chemical monopole and dipole moment of a product with a stoichiometric coefficient $\nu_i=1$.
	
	The solutes are treated within the Poisson-Boltmann-Debye-Hückel (PBDH) approximation,\cite{Burelbach2019} meaning that the local solute densities are described by the Poisson-Boltzmann distribution 
	\begin{equation}
	n_i(\mathbf r)=n_i^b\exp\left[-\frac{\phi_i(\mathbf r)}{k_BT}\right],\label{eq:-17}
	\end{equation}
	with $\left|\phi_i(\mathbf r)/(k_BT)\right| \ll 1$, where $\phi_i(\mathbf r)$ is the specific interaction potential of solute $i$ with the JP surface. Here, we assume this interaction to be purely electrostatic, in which case the interaction potential is given by $\phi_i(\mathbf r)=q_i\phi_E(\mathbf r)$,	where $\phi_E(\mathbf r)$ is the local electric potential within the interfacial layer. The gradient of $-\phi_E(\mathbf r)$ is not to be confounded with the electric field $\mathbf E(\mathbf r)$, which exclusively stems from the chemical activity of the JP. We further introduce the valency $z_i$ of a solute, such that $q_i=z_ie$, where $e$ is the elementary charge. Within the PBDH approximation, the linearised Poisson equation yields the well-known Yukawa form of the local electric potential if the surface charge distribution is isotropic. Here, we assume that this form remains valid for weak departures from this isotropy, such that
	\begin{equation}
	\phi_E(\mathbf r)=\zeta(\hat{\mathbf r})\frac{R}{r}\exp{-\kappa(r-R)},\label{eq:-18}
	\end{equation}
	where $\kappa=\left[\left(\sum_i n_i^bq_i^2\right)/(\epsilon k_BT)\right]^{1/2}$ is the inverse of the Debye screening length $\lambda_D$. The electric surface potential $\zeta(\hat{\mathbf r})$
	is related to the surface charge density of the colloid and may therefore be anisotropic. 
	
	By expanding eq.\ (\ref{eq:-17}) to first order in $\left|\phi_i(\mathbf r)/(k_BT)\right| \ll 1$ and using eq. (\ref{eq:-9}), the force density $\bm{\mathcal F}(\mathbf r)=\sum_{i\neq 0}n_i(\mathbf r)\mathbf F_i(\mathbf r)$ can be expressed as
	\begin{eqnarray}
	&&\bm{\mathcal F}(\mathbf r)=\label{eq:-19}\\
	&&\frac{3}{2}\eta\sum_{i\neq 0}\nu_iR_i\left(1-z_i\frac{e\phi_E(\mathbf r)}{k_BT}\right)\left[\frac{k\hat{\mathbf r}}{r^2}+\frac{1}{r^3}\left(3\hat {\mathbf r}\hat {\mathbf r}-\mathbf 1\right)\cdot\mathbf p\right],\nonumber
	\end{eqnarray}
	where we have also substituted $\xi_i=6\pi\eta R_i$.
	Using eq. (\ref{eq:-19}) to evaluate eq. (\ref{eq:-14}), the active angular velocity $\bm{\omega}_a$ finally takes the form
	\begin{equation}
	\bm{\omega}_a=-\frac{g(\kappa R)}{8R^2}\left\lbrace \sum_{i\neq 0}z_i\nu_iR_i\right\rbrace \left\langle\zeta'
	(\hat{\mathbf r})\hat{\mathbf r}\right\rangle_S\times\mathbf p,\label{eq:-22}
	\end{equation}
	where $\zeta'(\hat{\mathbf r})=e\zeta(\hat{\mathbf r})/(k_BT)$. The vector $\left\langle\zeta'(\hat{\mathbf r})\hat{\mathbf r}\right\rangle_S$ can thus be interpreted as the 'interfacial' dipole moment of the rescaled electric surface potential $\zeta'(\hat{\mathbf r})$. The dimensionless function $g(x)$ is always positive and given by
	\begin{equation}
	g(x)=-2+x(1-x)+(6+x^3)e^x E_1(x),\nonumber
	\end{equation}
	where $E_1(x)=\int_x^\infty t^{-1}e^{-t}dt$. 
	
	Hence, eq. (\ref{eq:-22}) only gives a non-zero angular velocity $\bm{\omega}_a$ if the chemical and interfacial dipole moments
	$\mathbf p$ and $\left\langle\zeta'(\hat{\mathbf r})\hat{\mathbf r}\right\rangle_S$ point in different directions. More generally, this shows that active rotation can only occur if the JP is not overall axisymmetric. Hence, if the chemical activity of the JP is axisymmetric, then active rotation requires an anisotropic interfacial solute layer that breaks this axisymmetry. This is the reason why active rotation is not observed for half-capped JPs.	
	
	\section{Conclusion}
	
	We have used the Onsager-Casimir reciprocal relations to describe the motion of self-diffusiophoretic JPs. Our approach is consistent with previous results and provides an extension of these results beyond the boundary-layer approximation. Moreover, identifying the electrochemical forces as thermodynamic forces within Onsager's theory has allowed us to naturally combine the effects of diffusio- and electrophoresis, showing that the active motion of a JP is completely determined by its chemical activity and specific interaction with the surrounding fluid.
	
	Although we have also made progress in the description of diffusiophoretic interactions, it must be noted that these results are only expected to apply to the far-field regime where the solvent maintains a hydrostatic equilibrium around the JPs. The description of such interactions remains a challenge in the near-field regime, where specific and hydrodynamic interactions between JPs become important and where mutual boundary conditions can no longer be ignored. An accurate reciprocal description of near-field diffusiophoretic interactions will therefore have to resort to more advanced ideas that remain to be explored.
	
	\section{Acknowledgements}
	
	JB gratefully acknowledges helpful discussions with Josua Grawitter.
	
	\bibliography{References}
	
\end{document}